\documentclass[iop,apj,numberedappendix]{emulateapj}

\usepackage{epstopdf}
\usepackage{natbib}
\usepackage{graphicx}
\usepackage{amsmath,amsfonts}
\slugcomment{Accepted for publication in ApJ, July 4, 2012}

\def\lsim{\mathrel{\raise.3ex\hbox{$<$\kern-.75em\lower1ex\hbox{$\sim$}}}}
\def\gsim{\mathrel{\raise.3ex\hbox{$>$\kern-.75em\lower1ex\hbox{$\sim$}}}}

\begin{document}
\title{Ultra-High Resolution Intensity Statistics of a Scintillating Source}
\shorttitle{Intensity Scintillation Statistics}

\author{M. D. Johnson, C. R. Gwinn}
\shortauthors{Johnson \& Gwinn}
\affil{Department of Physics, University of California, Santa Barbara, California 93106, USA}
\email{ michaeltdh@physics.ucsb.edu,cgwinn@physics.ucsb.edu}

\begin{abstract}
We derive the distribution of flux density of a compact source exhibiting strong diffractive scintillation. 
Our treatment accounts for arbitrary spectral averaging, spatially-extended source emission, and the possibility of intrinsic variability within the averaging time, as is typical for pulsars. 
We also derive the modulation index and present a technique for estimating the self-noise of the distribution, which can be used to identify amplitude variations on timescales shorter than the spectral accumulation time. Our results enable a for direct comparison with ultra-high resolution observations of pulsars, particularly single-pulse studies with Nyquist-limited resolution, and can be used to identify the spatial emission structure of individual pulses at a small fraction of the diffractive scale.
\end{abstract}


\section{Introduction}

\subsection{Noise and Scintillation}

Scintillation of compact astrophysical radio sources has proven to be a remarkable probe of both the turbulent interstellar plasma and the extreme conditions of the emission regions of these sources. 
The present work is primarily motivated by observations of pulsars at meter and decimeter wavelengths, for which the AU-scale scattering resembles a stochastic `lens' with a potential resolving power of less than a nanoarcsecond. However, our results apply to any compact object that exhibits diffractive scintillation.

The emission and scattering physics involve the interaction and mixing of many random variables. At the heart is the amplitude-modulated noise emitted by the source \citep{Rickett_AMN}. Indeed, even the amplitude modulation is stochastic, with individual pulses showing complex broadband profile structures, from bursty microstructure to strong and often non-Gaussian pulse-to-pulse variations \citep{Rickett_microstructure,Vela_lognormal,Kramer_Vela}. This emission then propagates through the dilute, turbulent plasma of the interstellar medium (ISM), where it undergoes dispersion and Faraday rotation, in addition to being strongly scattered. Finally, an observer samples the electric field in the presence of additive background noise, which often dominates the signal. Furthermore, the statistics of this random process evolve as the Earth, pulsar, and ISM move relative to each other.

Since the size of the scattering disk ($\Sigma$) scales approximately as the square of the observing wavelength ($\lambda$), the angular resolution afforded by the scintillation is $\theta_{\text{diff}} \sim \lambda / \Sigma \propto \lambda^{-1}$. Thus, longer wavelengths permit superior resolving power. However, at longer wavelength each scintillation element contains far fewer samples, because of the relatively smaller timescale ($\Delta t_{\mathrm{d}} \propto\! \lambda^{-1}$) and bandwidth ($\Delta \nu_{\mathrm{d}} \propto\! \lambda^{-4}$) of the scintillation pattern. It therefore becomes subtle to decouple the scintillation from the source and background noise. 

To address these difficulties, we present analytical expressions for the probability density function (PDF) of intensity after arbitrary temporal averaging of the spectra. We describe the case when single-pulse spectra are formed and then are averaged in groups of $N$ adjacent pulses; we exclusively use $N$ to denote such temporal averaging. We account for the decorrelation of the scintillation pattern during this averaging and derive the resulting modification to the intensity PDF. We also account for spatially-extended source emission and show that the leading order PDF correction is identical, up to scale, to that of the decorrelation of the scintillation pattern within the averaging time. In addition, we demonstrate that spatially-extended emission affects the measured PDF of intensity, even in the limit of $N=1$ averaged spectra. This feature presents the remarkable opportunity of measuring the emitting region sizes for individual pulses, or of subclasses of pulses. 

The sources and variety of noise are diverse, stemming from the background noise and the amplitude-modulated noise emitted by the source. 
Measured quantities such as intensity then correspond to random variables, and their stochastic variation introduces additional noise. 
Applied to the source, this noise is the familiar \emph{self-noise} \citep{Dicke,Kulkarni_SelfNoise,Zmuidzinas,Intermittent_Noiselike_Emission}, but the background and scintillation parameters have similar limitations.   
Such noise is generally unbiased and scales with intensity. 
Our derivations fully account for all these types of noise. We also account for artifacts arising from instrumental limitations, such as quantization of the analog signal and quadrature downconversion. These types of effects introduce additional noise as well as bias and distortion. 

To supplement our calculated PDFs, we also analyze expressions based on the moments of the distributions. In particular, we extend the modulation index, a traditional estimator of source emission geometry, to include the effects of self-noise and pulsar amplitude variations, and we analyze a technique that estimates self-noise. Our expressions refine these simple tools for intrinsic emission inference.

\subsection{Strategy for Comparison with Observations}

An observation well-suited for comparison with our models has an observing bandwidth $B \gg \Delta \nu_{\mathrm{d}}$, an observation duration $t_{\mathrm{obs}} \gg \Delta t_{\mathrm{d}}$, and a spectral accumulation time much longer than the pulse-broadening timescale. For each pulse period, the on-pulse and off-pulse regions yield estimates of the signal and noise intensities, respectively. 
These estimated parameters fully define a model of the intensity PDF (see \S\ref{sec::PDist}), which is readily compared with a histogram of the measured data. 

The residual structure in this comparison contains information about the spatial extent of the emission region and the evolution of the scintillation pattern, as well as artifacts from instrumental limitations and errors in the estimated on-pulse and off-pulse intensities. The on-pulse and off-pulse averages once again parametrize a model for this residual structure (see \S\ref{sec::Residual}), which is quantified by a single fitted parameter for amplitude. This fitted parameter yields the transverse size of the emission region, if the scattering geometry is specified.

\subsection{Comparison with Previous Work}

Most past work has focused on the regime of many samples of the spectrum averaged together: $N\rightarrow\infty$. 
For such data, the effects of noise can be approximated as Gaussian. 
\citet{ISO} gave the expected distribution of intensity for a strongly scintillating source, including the effects of a spatially-extended emission region.  
\citet{Gwinn00} extended these forms to account for Gaussian self-noise at small intensity, and estimated the contributions of averaging in time and frequency by deriving their effects on the modulation index.
In an unpublished manuscript, \citet{Cordes_Superresolution_1} presented expressions for the PDF of scintillation gain in terms of a Karhunen-Lo\`{e}ve expansion with the coefficients defined by a Fredholm equation, and suggested approximate forms determined by $\chi^2$ distributions with the appropriate number of degrees of freedom. 
\citet{IVSS} gave the distribution of interferometric visibility for a pointlike or extended source in strong scintillation.
\citet{Noise_Inventory} presented expressions for self-noise, suitable for large $N$.

The present work focuses on the complementary regime of single spectral realizations, and averages of $N$ such spectra. We connect with these previous works through the asymptotic forms of our distributions. Similarly, \citet{Van_Straten_Polarimetry} analyzed statistics of Stokes parameters when $N=1$ and showed, for instance, that the degree of polarization is undefined in this case. His results describe the high signal-to-noise regime with no effects of propagation or source emission structure, as appropriate for the high-frequency observations of giant pulses from the Crab pulsar that motivated his work \citep{Hankins_Crab}. 

\subsection{Outline of Paper}

In \S\ref{sec::Theory}, we briefly review the theoretical descriptions of pulsar emission and interstellar scintillation, and establish our assumptions about each. 
Then, in \S\ref{sec::PDist}, we present an expression for the exact intensity PDF of a scintillating point source, as well as useful approximations. 
We next apply these approximations in \S\ref{sec::Residual} to estimate modifications to the PDF arising from the decorrelation of the scintillation pattern within the averaging time, effects of finite source emission size, and observational constraints such as finite scintillation averages; we then use these expressions to derive the limiting resolution for emission size inference.
In \S\ref{sec::Moments}, we derive the modulation index and analyze a technique that identifies self-noise. Finally, in \S\ref{sec::Summary}, we summarize our results and outline the prospects for pulsar observations.

\section{Theoretical Background}
\label{sec::Theory}

\subsection{Pulsar Radio Emission}

Pulsar radio emission is thought to arise from coherent curvature radiation from a relativistic electron-positron plasma, flowing outward along open field lines from the stellar surface to the light cylinder. 
Charge acceleration occurs in a `gap' of depleted charge density where the force-free state cannot be maintained. Such sites initiate a pair-production cascade, which forms a secondary plasma \citep{Sturrock,Ruderman_Sutherland}. This secondary plasma generates the radio emission, which is then beamed, ducted, and refracted as it propagates through the upper magnetosphere \citep{Barnard_Arons,Arons_Barnard,Lyutikov_Parikh}. The gap may be located close to the polar cap region (the `polar gap') \citep{Arons_Scharlemann}, or along the boundary of the open and closed field lines (the `slot gap') \citep{Arons_SlotGap}. An `outer gap' \citep{Cheng_OuterGap,Romani_OuterGap}, situated close to the light cylinder radius, is the favored site for much high-energy emission and may also contribute to the radio emission. Differing locations of the gap may be responsible for some of the emission variety between pulsars.

The superposition of many independent radiators produces white Gaussian noise. 
Thus, although the radio emission process is fundamentally coherent, by groups of many particles, the bulk emission may still be effectively spatially and temporally incoherent. 
Intrinsic physical processes (and the pulsar's rotation) then modulate the envelope of this random field, motivating an `amplitude-modulated noise' description of the emission \citep{Rickett_AMN}. Our results assume this type of emission, confined to a region that is a small fraction of the diffractive scale 
$r_{\mathrm{d}} \equiv \lambda D/(2\pi \Sigma) \sim 10^6 \mathrm{\ m}$, where $D$ is the characteristic Earth-scatterer distance.   

\subsection{Interstellar Scattering}
\label{sec::ISS_Scattering}
Density inhomogeneities in the dilute, turbulent plasma of the ISM scatter the radio emission, leading to multipath propagation. For most pulsars observed at decimeter wavelengths, the phases between pairs of paths differ by many turns and are therefore uncorrelated; this is the regime of `strong' scattering. Equivalently, the diffractive scale is much smaller then the Fresnel scale, $r_{\mathrm{F}} \equiv 
\sqrt{\lambda D} \sim 10^9 \rm{\ m}$.

In general, the timescale for the evolution of the diffraction pattern at the observer ranges from seconds to hours, and so is effectively static over the scattering timescale.
In this case, the effect of scattering is to convolve the pulsar signal with a `propagation kernel.' The characteristic width of the average propagation kernel is the pulse-broadening timescale $t_0$. Dispersion contributes an additional frequency-dependent delay that is readily reversed \citep{Hankins_Coherent}. The motivation and limitations of approximating scattering as a convolution are given in much greater detail by \citet{Intermittent_Noiselike_Emission}.

Although this picture of strong scattering of a point source is quite general, a description of the effects of a spatially-extended emission region requires specification of the scattering geometry. We present explicit results for the case of thin-screen scattering, as is observationally motivated \citep{Williamson,Komesaroff72,Hill_05}.  
We give additional reductions by assuming a square-law phase structure function; the generalization to Kolmogorov or otherwise is straightforward and changes only the scale of the PDF modifications that we derive.

\subsection{Instrumental and Observational Requirements}
\label{sec::instrumental}
In practice, spectra must be formed over a finite accumulation time $t_{\mathrm{acc}}$; the convolution representation of scattering requires $t_{\mathrm{acc}} \gg t_0$ (see \citet{Intermittent_Noiselike_Emission} for additional details). The unique nature of pulsars and their generally short duty cycles allow a particularly elegant limiting option: the formation of single-pulse spectra that contain \emph{all} pulsed power. Our descriptions are exact in this limit, and appropriate whenever $t_{\mathrm{acc}} \gg t_0$.

We assume that such spectra are formed and then averaged in groups of $N$ adjacent pulses. We also assume that the averaging timescale is much shorter than the scintillation timescale -- the `snapshot image' of \citet{NarayanGoodman89} -- which subsumes the much weaker condition that the scattering material is approximately static during the accumulation of a single spectrum.

Frequency averaging is analogous but inherits additional information from non-stationary signals. The intrinsic pulse shape incurs spectral mixing via its associated frequency-domain convolution; the average is then of correlated intensities (see \S\ref{sec::ConvolutionCorrelated}). Thus, a description of frequency averaging requires the specification of individual pulse profiles, whereas a description of temporal averaging requires only the phase-averaged source intensity for each pulse.

Lastly, we assume that the data explore a large representation of the full ensemble of diffractive scintillation: 
$(B /\Delta \nu_{\mathrm{d}} )\! \times\! (t_{\mathrm{obs}}/\Delta t_{\mathrm{d}}) \gg 1$.  
Comparison with observations is most powerful if the stronger condition $B \gg \Delta \nu_{\mathrm{d}}$ is satisfied so that the intrinsic amplitudes of individual pulses can be estimated.

\subsection{Notation}
Despite the number and variety of random variables involved in the emission and scattering processes, much statistical homogeneity exists; our notation highlights this foundational uniformity. We use $z_x$ to denote a circular complex Gaussian random variable with unit variance, indexed by $x$. Some common examples of such variates are a single realization of the scalar electric field or the complex gain from propagation.
The PDF of $z_x$ is exponential in its squared norm, $P(z_x) = \pi^{-1} e^{-|z_x|^2}\!$, with respect to the standard complex metric $d\mathrm{Re}[z_x] d\mathrm{Im}[z_x]$. This norm is implicit for all PDFs presented with respect to complex random variables.

We also encounter many exponential random variables, usually in connection with a scintillation ``gain,'' which are thus denoted $G$ or $G_i$, with $P(G) = e^{-G}$ for $G>0$.
Note that we use $P()$ to generically denote a PDF with respect to the given variables and parameters, as in the previous examples.

\section{Distribution of Intensity}
\label{sec::PDist}

We now derive the expected PDF of intensity when the spectra are averaged over groups of $N$ consecutive pulses. We assume that the scintillation pattern is frozen during each average, and that the source emission is pointlike.
In \S\ref{sec::Residual}, we relax these assumptions and discuss practical limitations. 
Our results apply to scalar electric fields that have been baseband shifted and coherently dedispersed.

We first derive an exact expression for the PDF of intensity in \S\ref{sec::PDist_I}. However, for $N>1$ this expression has a removable singularity at the origin and is analytically burdensome. These features motivate various alternatives, such as the \emph{i.i.d.\ approximation} and the \emph{gamma approximation}, which we derive in the following \S\ref{sec::iid_Approximation} and  \S\ref{sec::Gamma_Approximation}, respectively. We present a traditional representation, the \emph{Gaussian approximation}, in \S\ref{sec::GaussianApproximation} for comparison. Both the i.i.d.\ and gamma approximations are simple and powerful, and we heavily utilize them for the remainder of the paper.

\subsection{The Exact Intensity PDF}
\label{sec::PDist_I}

The pulsar emits white Gaussian noise $\epsilon_i$, modulated by a time-varying envelope $\sqrt{I_{\mathrm{s}}} f_i$. 
Here, $I_{\mathrm{s}}$ is a constant characteristic scale of the source intensity, $\epsilon_i$ is normalized to have unit variance, and the modulation $f_i$ is power-preserving. To account for pulse-to-pulse variations in the phase-averaged flux density, we also include a dimensionless pulse amplitude factor $A_j$, where $j$ indexes the spectrum (i.e. the pulse number). We require no assumptions about the timescale of variability of the envelope $f_i$ or the nature of the pulse-to-pulse variations $A_j$. Our treatment therefore accommodates arbitrary variability of the pulsar, such as the possibility of correlated pulse-to-pulse variations, log-normal amplitude statistics, or nanosecond-scale bursts, which merely modulates the Gaussian self-noise.

The propagation then acts to convolve this intrinsic emission with a scattering kernel, as discussed in \S\ref{sec::ISS_Scattering}. 
The kernel similarly consists of a power-preserving envelope $g_i$ that modulates white Gaussian noise $\eta_i$ of unit variance. The propagated signal is then superimposed with the background noise $\sqrt{I_{\mathrm{n}}}\beta_i$, where $I_{\mathrm{n}}$ is the amplitude of the noise, and $\beta_i$ is white Gaussian noise of unit variance. We assume that the background noise level is effectively constant over the averaging time, although this restriction is easily relaxed.  

The observed scalar electric-field time series $x_i$ and its Fourier-conjugate spectrum are thus given by
\begin{align}
\label{eq::E_I}
x_i &= \sqrt{A_j I_{\mathrm{s}}} \left[ \left( f \epsilon \right) \ast \left(g \eta \right) \right]_i + \sqrt{I_{\mathrm{n}}}\beta_i \\
\nonumber \Rightarrow \tilde{x}_i &= \sqrt{A_j I_{\mathrm{s}}} \left( \tilde{f} \ast \tilde{\epsilon} \right)_i \left(\tilde{g} \ast \tilde{\eta} \right)_i + \sqrt{I_{\mathrm{n}}} \tilde{\beta}_i. 
\end{align}
Here, we denote Fourier conjugate variables with a tilde. In short, this equation describes the amplitude-modulated noise of the pulsar ($\sqrt{I_{\mathrm{s}}} f_i\epsilon_i$) convolved with the stochastic propagation kernel ($g_i \eta_i$) and added to background noise ($\sqrt{I_{\mathrm{n}}} \beta_i$). Conventionally, these terms are labeled by their noise-free limits: the pulse profile $|f_i|^2\!$, pulse-broadening function $|g_i|^2\!$, gated signal $I_{\mathrm{s}}$, and background $I_{\mathrm{n}}$.

Of course, $\tilde{\epsilon}_i$, $\tilde{\eta}_i$, and $\tilde{\beta}_i$ are mutually independent (circular complex Gaussian) white noise. We see that a single spectral sample $\tilde{x}_i$ is of the form $\sqrt{A_j I_{\mathrm{s}}} z_{\mathrm{f}} z_{\mathrm{g}} + \sqrt{I_{\mathrm{n}}} z_{\mathrm{b}}$, where $z_{\mathrm{f}}$, $z_{\mathrm{g}}$, and $z_{\mathrm{b}}$ are each circular complex Gaussian random variables with unit variance. If the scintillation is held fixed (e.g.\ $z_{\mathrm{g}} = \mathrm{const.}$), then the intensity $|\tilde{x}_i|^2$ is drawn from an exponential distribution with scale $\bar{I}_j \equiv A_j I_{\mathrm{s}} |z_{\mathrm{g}}|^2 + I_{\mathrm{n}}$. 
The PDF of the average of $N$ such intensities is then given by Eq.\ \ref{eq::mult_I}:
\begin{align}
\label{eq::I_G_PDF}
P(I;N | z_{\mathrm{g}}) &= N \sum_{j=1}^N \left( \frac{\bar{I}_j^{N-2} }{\prod\limits_{\substack{\ell = 1\\ \ell\neq j}}^N \left(\bar{I}_j - \bar{I}_\ell \right)}  \right) e^{-N I/\bar{I}_j}\\
\nonumber &= \frac{N}{ \left( I_{\mathrm{s}} |z_{\mathrm{g}}|^2 \right)^{N-1} } \\
\nonumber & \quad {} \times \sum_{j=1}^N  \frac{\left(A_j I_{\mathrm{s}} |z_{\mathrm{g}}|^2  + I_{\mathrm{n}} \right)^{N-2} }{\prod\limits_{\substack{\ell = 1\\ \ell\neq j}}^N  \left(A_j - A_\ell \right)} e^{-\frac{N I}{\left(A_j I_{\mathrm{s}} |z_{\mathrm{g}}|^2 + I_{\mathrm{n}} \right)}}.
\end{align}
Observe that when we designate $N$ as a distribution parameter, we implicitly include the background $I_{\mathrm{n}}$ and source amplitudes $\{ A_j I_{\mathrm{s}} \}$.

\begin{figure}[t]
\includegraphics*[width=0.48\textwidth]{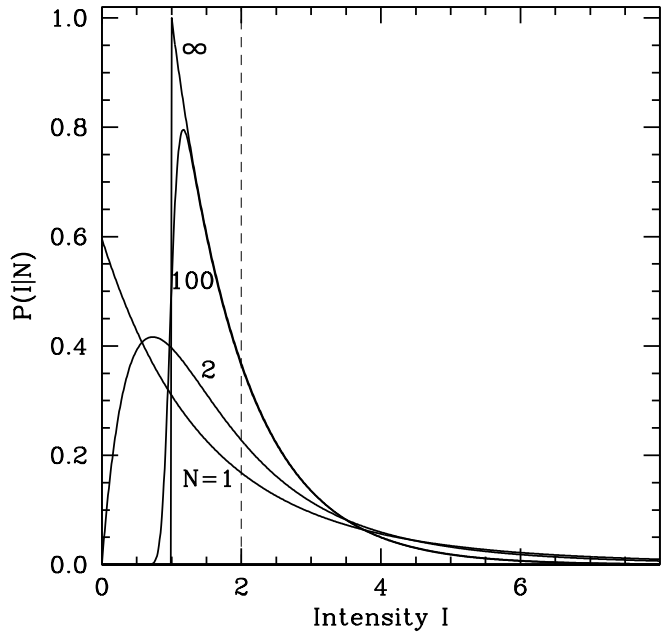}
\caption{
PDF of intensity for different pulse averaging numbers $N\!=\!1$, 2, 100, and $\infty$. The source and background intensities are unity, as are all the pulse amplitude factors $A_j$; the dashed line denotes the mean of $2.0$. Because of the pulse amplitude degeneracy, the i.i.d.\ representation (Eq.\ \ref{eq::I_approx}) must be used (and is exact).
}

\label{fig_pdf_iid}
\end{figure}

All that remains is to incorporate the distribution of the scintillation random variable $z_{\mathrm{g}}$. We assume that $z_{\mathrm{g}}$ explores a representation of the full diffractive ensemble. We introduce a `scintillation gain' parameter $G \equiv |z_{\mathrm{g}}|^2$, so $P(G) = e^{-G}$. Integrating the scintillation ensemble through $G$ then provides the PDF of intensity:
\begin{align}
\label{I_PDF}
P(I;N) = &N \int_0^\infty dG \ \frac{e^{-G}}{ \left(I_{\mathrm{s}} G\right)^{N-1} } \\
\nonumber & \qquad {} \times \sum_{j=1}^N  \frac{\left(A_j I_{\mathrm{s}} G  + I_{\mathrm{n}} \right)^{N-2} }{\prod\limits_{\substack{\ell = 1\\ \ell\neq j}}^N  \left(A_j - A_\ell \right)} e^{-\frac{N I}{\left(A_j I_{\mathrm{s}} G + I_{\mathrm{n}} \right)}}.
\end{align}
This equation gives the expected PDF of intensity for the average of $N$ pulses with amplitudes $\{ I_{\mathrm{s}} A_i \}$ and fixed scintillation, in the presence of white background noise with mean intensity $I_{\mathrm{n}}$, if the spectra explore a representative ensemble of diffractive scintillation. Examples of this distribution for various $N$ are shown in Figure \ref{fig_pdf_iid}.

Numerical evaluation of the PDF can be challenging because of the singularity at $G = 0$, although the integrand is not divergent. To address this difficulty, we employ variable-precision arithmetic libraries; the additional precision is essential for $N \gsim 10$.

\subsection{Approximating the Intensity PDF}
\label{I_PDF_iid}

Although we have derived the exact expression for the intensity PDF (Eq.\ \ref{I_PDF}), we now derive several useful approximations. These approximations resolve the singularity at $G = 0$, uncover the $N\rightarrow \infty$ behavior, and present a simplified framework for deriving and interpreting perturbations of the PDF arising from effects such as a finite size of the emission region, decorrelation of the scintillation pattern within the averaging time, and errors in model parameters.

\subsubsection{The i.i.d.\ Approximation}
\label{sec::iid_Approximation}
We derive our first approximation by replacing each set of $N$ pulse amplitude factors $A_j$ by its average $A \equiv \langle A_j \rangle$. This replacement simplifies the statistics because the intensities that are averaged under such assumptions are independent and identically distributed (i.i.d.); thus, we refer to this approximation scheme as the \emph{i.i.d.\ approximation}.

We emphasize that the i.i.d.\ approximation does not ignore amplitude variability of the pulsar, but does reduce the variability to a moving average. The difference is critical because, for example, the i.i.d.\ approximation is exact when $N=1$.

After substituting the local amplitude average $A$, the PDF of intensity during a fixed scintillation gain $G$ (i.e.\ Eq.\ \ref{eq::I_G_PDF}) is given by an Erlang distribution (Eq.\ \ref{eq::I_iid}):
\begin{align}
\label{eq::I_approx}
P(I;N|G) = \frac{N^N}{(N-1)!} \frac{I^{N-1}}{\left(A I_{\mathrm{s}} G + I_{\mathrm{n}}\right)^N}  e^{-\frac{N I}{A I_{\mathrm{s}} G + I_{\mathrm{n}}}}.   
\end{align}
As for Eq.\ \ref{I_PDF}, the scintillation ensemble can then be included via its exponential density: $P(I;N) = \int_0^\infty dG\ e^{-G} P(I;N|G)$.

\subsubsection{The Gamma Approximation}
\label{sec::Gamma_Approximation}

We now extend the i.i.d.\ approximation to partially account for the effects of variation in the pulse amplitudes $A_j$ within sets of $N$ pulses. 
To achieve this extension, we replace $N$ in Eq.\ \ref{eq::I_approx} by an effective number of degrees of freedom $N_{\mathrm{eff}}$, chosen such that the mean and variance match those of the exact PDF. Because $N_{\mathrm{eff}}$ is not necessarily an integer, the factorial in the Erlang distribution must be replaced by a gamma function; this modified form is the gamma distribution. 
The mean and variance of this gamma distribution are
\begin{align}
\label{eq::moments_gamma}
\mu_{\rm gamma} &= A I_{\mathrm{s}} G + I_{\mathrm{n}},\\
\nonumber \sigma_{\rm gamma}^2 &= \frac{(A I_{\mathrm{s}} G + I_{\mathrm{n}})^2}{N_{\mathrm{eff}}}.
\end{align}
For the exact distribution, given by Eq.\ \ref{eq::I_G_PDF}, the mean is simply the average of the marginal means, and the variance is the average of the marginal variances, divided by $N$:
\begin{align}
\label{eq::moments_exact}
\mu_{\rm exact} &= \frac{1}{N}\sum_{j=1}^N \left( A_j I_{\mathrm{s}} G + I_{\mathrm{n}}\right), \\
\nonumber \sigma_{\rm exact}^2 &= \frac{1}{N^2}\sum_{j=1}^N  \left( A_j I_{\mathrm{s}} G + I_{\mathrm{n}}\right)^2.
\end{align}
Matching Eq.\ \ref{eq::moments_gamma} and Eq.\ \ref{eq::moments_exact} gives the correspondence appropriate for the gamma distribution:
\begin{align}
A &= \frac{1}{N} \sum_j A_j, \\
\nonumber N_{\mathrm{eff}} &= \frac{\left(\sum_{j=1}^N 1 + A_j I_{\mathrm{s}} G/I_{\mathrm{n}} \right)^2}{ \sum_{j=1}^N \left(1 + A_j I_{\mathrm{s}} G/I_{\mathrm{n}} \right)^2} \leq N.
\end{align}
The scintillation ensemble is again included via integration over $G$ with its exponential weight.

We refer to this extension of the i.i.d.\ approximation as the \emph{gamma approximation}. The gamma approximation partially accounts for the pulse-to-pulse variability, and substantially improves the i.i.d.\ approximation for $N \gg 1$.

\subsubsection{The Gaussian Approximation}
\label{sec::GaussianApproximation}
By the Central Limit Theorem, Eq.\ \ref{eq::I_approx} approaches a Gaussian distribution as $N \rightarrow \infty$,  with mean $\bar{I} = A I_{\mathrm{s}} G + I_{\mathrm{n}}$ and variance 
$\delta I^2 = \left(A I_{\mathrm{s}} G + I_{\mathrm{n}} \right)^2\!/N = \bar{I}^2/N $.
Accounting for the amplitude variations that are ignored by the i.i.d.\ approximation gives the generalized form, $\delta I^2 = \left[ \bar{I}^2 + 2 (\bar{I}-I_{\mathrm{n}})^2 \frac{\delta A^2}{\langle A \rangle^2} \right]/N$; i.e.\ pulse-to-pulse amplitude variations contribute twice their variance to the source self-noise. These relationships both reflect the familiar fact that self-noise is inversely proportional to the number of averaged samples $N$ \citep{Dicke}. 

A traditional approximation strategy replaces $P(I;N|G)$ with this matched Gaussian distribution; we refer to this scheme as the \emph{Gaussian approximation}.

\subsection{Utility of Approximations}
These three approximations are particularly useful because they reduce the $N+1$ degrees of freedom in a single averaged spectra to three: an overall scale partitioned into source and background contributions, and a measure of the averaging. In fact, the i.i.d.\ approximation has only two degrees of freedom because the degree of averaging is $N$; this reduction comes at the expense of ignoring pulse amplitude variance.

\begin{figure}[t]
\includegraphics*[width=0.48\textwidth]{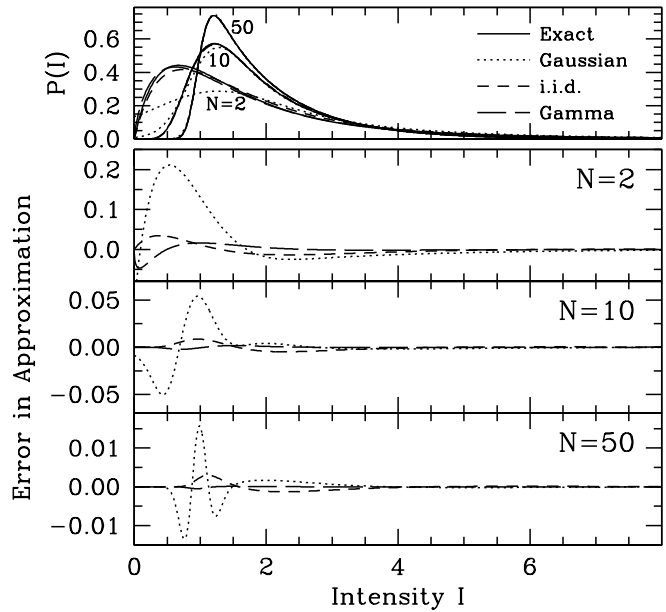}
\caption{
Exact PDF of intensity for $N\!=\!2$, 10, and 50, and the three approximations discussed in \S\ref{I_PDF_iid}. The source and background intensities are unity, and the $N$ pulse amplitude factors are uniformly distributed between $0$ and $2$.
}
\label{fig_approx_compare}
\end{figure}

Figure \ref{fig_approx_compare} demonstrates the relative quality of these three approximations in the presence of substantial amplitude variations.
The Gaussian approximation is inferior to both the i.i.d.\ and gamma approximations, particularly for small $N$; indeed, the Gaussian approximation includes nonzero probability for negative intensities. The i.i.d.\ and gamma approximations, on the other hand, give excellent representations for arbitrary $N$ and are much easier to implement numerically than the exact representation of Eq.\ \ref{I_PDF}, particularly for large $N$. 

Our principal use of these approximations is to derive analytical estimates for PDF modifications arising from effects such as finite emission size and decorrelation of the scintillation pattern within the averaging time. We favor the relatively simple i.i.d.\ approximation for this purpose. While the small errors in the approximations might be significant in some measured PDFs, we expect the similarly small errors on the expected $\lsim$$1\%$ PDF modifications to be negligible. Furthermore, the most intriguing degree of averaging is $N=1$, which allows single-pulse studies and complete decoupling of emission size from decorrelation; both the i.i.d.\ and gamma approximations are, of course, exact for this case.

We therefore recommend the exact representation for PDF comparisons (or the gamma approximation if its accuracy suffices), and now derive PDF modifications using the i.i.d.\ approximation.

\section{PDF Modifications}
\label{sec::Residual}

We now analyze effects that modify the PDF of intensity relative to the form derived in \S\ref{sec::PDist}. Modifications can arise from physical processes both intrinsic and extrinsic to the source, as well as from instrumental effects. 
Since the model PDF of intensity of \S\ref{sec::PDist} depends only on direct observables (the average on-pulse and off-pulse intensities), the residual structure in a measured distribution relative to the corresponding model reflects these modifications. We also analyze the consequences of errors in the model parameters. 

We find degeneracies among these many types of effects. 
Many of these degeneracies can be resolved by varying the degree of averaging $N$ or the bandwidth used to construct the PDFs. 
We quantify the significance of each effect in terms of its expected bias on the inferred emission size $\gamma_{\mathrm{s}}$ and give strategies for identifying each source of bias.

However, despite the nearly identical form resulting from these varied effects, the derivation for each is unique. 
First, we account for decorrelation of the scintillation pattern within the averaging time in \S\ref{sec::Decorr}. Then, we derive the effects of spatially-extended source emission in \S\ref{sec::Size}. These two effects are degenerate, to excellent approximation, and we analyze their relative strength in \S\ref{sec::SizeDecorrCompare}. Next, we calculate the consequences of both noise and bias in the model parameters in \S\ref{sec::Obs}, and we estimate the influence of instrumental distortion in \S\ref{sec::saturation}. Finally, we derive the expected spatial resolution afforded by these scintillation statistics in \S\ref{sec::Expected_Resolution}.

\subsection{Effects of Temporal Decorrelation}
\label{sec::Decorr}

The scintillation pattern slowly evolves in response to the relative motions of the Earth, pulsar, and scattering medium, thereby affecting statistics of averaged intensities. We now derive the effects of this evolution on the observed distribution of intensity, in terms of the temporal autocorrelation function of the observed scintillation pattern. 

Although the scintillation pattern evolves somewhat similarly with frequency, the effects of frequency averaging fundamentally differ from those of temporal averaging.
Namely, for temporal averaging, the intensities within each average are independent, with correlated scales (i.e.\ the respective scintillation gains); whereas for frequency averaging, both the intensities within each average \emph{and} their scales can be correlated because of the intrinsic amplitude modulation. As noted in \S\ref{sec::instrumental}, spectral statistics after frequency averaging depend on individual pulse profiles, whereas spectral statistics after temporal averaging only depend on the phase-averaged flux densities.

Procedurally, we must index the scintillation factors $G$ by pulse number. The resulting set $\{ G_i \}$ is then a collection of correlated exponential random variables, with covariance determined by the autocorrelation function of the measured dynamic spectrum:
\begin{align}
\Gamma_{ij} &\equiv \left \langle G_i G_j \right \rangle -1\\
\nonumber &= \frac{\left \langle \left[ I(t,\nu) - I_{\mathrm{n}} \right]\left[ I(t+\Delta t_{ij},\nu) - I_{\mathrm{n}} \right] \right \rangle}{\left \langle I(t,\nu) - I_{\mathrm{n}} \right \rangle^2} - 1.
\end{align}
Details of this type of multivariate exponential distribution are given in Appendix \ref{sec::multI}.

In the `snapshot image' regime, the averaging timescale is much shorter than the decorrelation timescale of the scintillation pattern, so $\Gamma_{ij} \approx 1$. The leading order decorrelation effects in $(1-\Gamma_{ij})$ then provide an excellent approximation. Using Eq.\ \ref{eq::multI_approx}, we obtain
\begin{align}
\label{eq::Decorr_1}
&P(I;N,\{\Gamma_{ij}\}) - P(I;N)\\
\nonumber &\ \approx  {} - \sum_{i<j} \left(1 - \Gamma_{ij} \right) \int_0^\infty dG\ G e^{-G} \left.\frac{\partial^2 P(I;N|\{G_i\})}{\partial G_i \partial G_j}\right\rfloor_{G}\!.
\end{align}
Note that the partial derivatives are easily evaluated using the characteristic function of $P(I;N|\{G_i\})$: $\varphi(k) = \prod_\ell \left(1-i \bar{I}_\ell k/N\right)^{-1}$, where $\bar{I}_\ell \equiv A_j I_{\mathrm{s}} G_j + I_{\mathrm{n}}$. For example, applying the i.i.d.\ approximation ($A_j \equiv A$) yields the following relationship:
\begin{align}
\label{eq::Decorr_iid}
\left.\frac{\partial^2 P(I;N|\{G_i\})}{\partial G_i \partial G_j}\right\rfloor_{G} = \frac{1}{N(N+1)} \frac{\partial^2 P(I;N|G)}{\partial G^2}.
\end{align}

We can also reduce the sum over $\Gamma_{ij}$ in Eq.\ \ref{eq::Decorr_1} by approximating the evolution of the scintillation pattern as relative motion of the pulsar and observer with respect to a `frozen' random screen. The decorrelation in time is then simply expressed in terms of the phase structure function of the scattering medium \citep{Goodman_SO,CAN_89}. 
For example, a square-law phase structure function has $\Gamma_{ij} = \exp\left[ - \left(\Delta t_{ij}/\Delta t_{\mathrm{d}} \right)^2 \right]$, even in the presence of axial anisotropy \citep{Rickett77,GUPPI_Scattering}. If the averaging timescale is much less than the decorrelation timescale, then $\sum_{i<j}\left( 1 - \Gamma_{ij} \right) \approx N^2 \left(N^2 - 1\right)/\left(12 \Delta \tau_{\mathrm{d}}^2\right)$, where $\Delta \tau_{\mathrm{d}}$ is the decorrelation timescale in pulse periods. The quartic dependence on $N$ leads to a rapid onset of decorrelation artifacts in the observed PDF after only modest averaging. 

Combining this approximation for the decorrelation behavior with the general form for the PDF modification (Eq. \ref{eq::Decorr_1}), and applying the reduction afforded by the i.i.d.\ approximation (Eq. \ref{eq::Decorr_iid}) then provides the following estimate:
\begin{align}
\label{eq::Decorr_1_iid}
&P(I;N,\{\Gamma_{ij}\}) - P(I;N) \\
\nonumber & \qquad \approx {} - \frac{N(N-1)}{12 \Delta \tau_{\mathrm{d}}^2} \int_0^\infty dG\ G e^{-G} \frac{\partial^2 P(I;N|G)}{\partial G^2}.
\end{align}

\subsection{Effects of Extended Source Emission}
\label{sec::Size}

A spatially-extended emission region smoothes the fluctuations from scintillation by superimposing many slightly offset copies of the diffraction pattern at the observer. In fact, extended source emission is merely one example in a broad class of physical effects that alter the distribution of scintillation gain. Other possibilities arise from modified scattering assumptions, such as weak scintillation or L\'{e}vy-type scattering statistics \citep{LevyFlights}. To account for these cases, we introduce a generalized scintillation gain random variable $\mathcal{G}$, which is no longer constrained to follow exponential statistics. Specification of $P(\mathcal{G})$, combined with $P(I;N|\mathcal{G})$ given by Eq.\ \ref{I_PDF}, then provides the intensity PDF $P(I;N)$.

\citet{ISO} examined $P(\mathcal{G})$ for thin-screen scattering of a Gaussian distribution of source intensity. They demonstrated that, if the source emission region is small relative to the magnified diffractive scale, $\mathcal{G}$ is well-approximated by the convolution of three independent exponential random variables. 
The scales of these random variables reflect the size of the emission region. \citet{IVSS} derived an analogous result for interferometric visibility. 

By extending the description of  \S\ref{sec::PDist_I}, we now demonstrate that an equivalent form arises for any spatially-incoherent, compact emission region in strong scattering. We parametrize the transverse coordinates of the source by $\textbf{s}$. Each emitting location has independent source noise, and the emission envelope at each location may also vary. However, the propagation kernel is correlated over the emission region. A single spectral sample thus takes the form
\begin{align}
\label{eq::size_1}
\tilde{x}_i = \left\{ \int d^2 \textbf{s} \ \sqrt{A(\textbf{s}) I_{\mathrm{s}}} z_{\mathrm{f}}(\textbf{s}) z_{\mathrm{g}}(\textbf{s}) \right\} + \sqrt{I_{\mathrm{n}}} z_{\mathrm{b}}.
\end{align}
We specify our source coordinates, $\mathbf{s}$, to be centered on the spatial mean of source intensity: $\int d^2\textbf{s}\ \textbf{s} A(\mathbf{s}) = \mathbf{0}$. 

If the emission is confined to a small fraction of the magnified diffractive scale, then the scintillation random variable, $z_{\mathrm{g}}(\textbf{s})$, will be strongly correlated over the region of source intensity. We therefore expand to linear order: $z_{\mathrm{g}}(\textbf{s}) \approx z_{\mathrm{g}}(\textbf{0}) + \left. (\textbf{s} \cdot \nabla)  z_{\mathrm{g}} \right \rfloor_{s=0}$. 
The source term in Eq.\ \ref{eq::size_1} is then a convolution of three complex Gaussian random variables. Because of our choice of origin for $\textbf{s}$, these three random variables are mutually uncorrelated during a fixed scintillation pattern. The marginal variances therefore add to give $\mathcal{G}$ (up to overall normalization). In addition, the scintillation random variable, $z_{\rm g}(\textbf{s})$, is uncorrelated with its spatial derivatives, so the scales of the three variances are mutually independent. Thus, $\mathcal{G}$ is the convolution of three independent exponential random variables:
\begin{align}
\mathcal{G} \equiv \frac{1}{1 + \gamma_{\mathrm{s},1} + \gamma_{\mathrm{s},2}} \left( G + \gamma_{\mathrm{s},1} G_1 + \gamma_{\mathrm{s},2} G_2 \right).
\end{align}
The dimensionless subsidiary scales $\gamma_{\mathrm{s},i} \ll 1$ contain information about the transverse extent of the source emission. 
The scaling prefactor is merely chosen so that $\langle \mathcal{G} \rangle = 1$ and could equally well be absorbed into the definition of source intensity. 

Using the results of \citet{ISO}, we now explicitly relate the parameters $\gamma_{\mathrm{s},i}$ to the emission geometry by assuming a Gaussian distribution of source intensity and thin-screen scattering. In this case, the $\gamma_{\mathrm{s},i}$ give the squared size of the source in orthogonal directions $\hat{x}_i$, in units of the magnified diffractive scale:
\begin{align}
\label{eq::gamma_s}
\gamma_{\mathrm{s},i} = \left( \frac{D}{R} \frac{\sigma_i}{\frac{1}{2\pi}\frac{\lambda}{\theta_i} } \right)^2\!.
\end{align}
Here, $D$ is the characteristic observer-scatterer distance, $R$ is the characteristic source-scatterer distance, $\lambda$ is the observing wavelength, $\theta_i$ is the angular size of the scattering disk along $\hat{x}_i$, and $\sigma_i$ is the standard deviation of the distribution of source intensity along $\hat{x}_i$. 
This representation accommodates anisotropy of both the source emission and the scattering. 
If the source emission arises from a circular Gaussian intensity profile, then the FWHM of the emission region is $2\sqrt{\ln 4} \sigma \approx 2.35\sigma$, where $\sigma \equiv \sigma_i$.

Efforts to infer emission size often attempt to isolate the distribution $P(\mathcal{G})$ through sufficient spectral averaging; the modulation index $m$ then quantifies the effects of emission size on this distribution \citep{Salpeter67,Cohen67,ISO}:
\begin{align} 
\label{eq::modulation_index_simple}
m^2 &= \frac{\langle \mathcal{G}^2 \rangle}{\langle \mathcal{G} \rangle^2} - 1\\
\nonumber &= \frac{1 + \gamma_{\mathrm{s},1}^2 + \gamma_{\mathrm{s},2}^2}{\left( 1 + \gamma_{\mathrm{s},1} + \gamma_{\mathrm{s},2} \right)^2} \approx 1 - 2\left( \gamma_{\mathrm{s},1} + \gamma_{\mathrm{s},2} \right).
\end{align}
In \S\ref{sec::Modulation_Index}, we discuss the modulation index in depth and extend Eq.\ \ref{eq::modulation_index_simple} to account for the effects of self-noise, temporal decorrelation within the averaging time, and pulsar amplitude variability.

To determine the PDF modification resulting from a finite emission size, we first expand the PDF for a fixed scintillation pattern to linear order in $\gamma_{\mathrm{s},i}$:
\begin{align}
&P(I;N,\gamma_{\mathrm{s}}|\mathcal{G}) - P(I;N,\gamma_{\mathrm{s}}{=}0|G)\\ 
\nonumber & \approx \left[ G_1 \gamma_{\mathrm{s},1} + G_2 \gamma_{\mathrm{s},2} - G\left( \gamma_{\mathrm{s},1} + \gamma_{\mathrm{s},2} \right) \right] \frac{\partial P(I;N,\gamma_{\mathrm{s}}{=}0|G)}{\partial G} .
\end{align}

\begin{figure}[t]
\includegraphics*[width=0.48\textwidth]{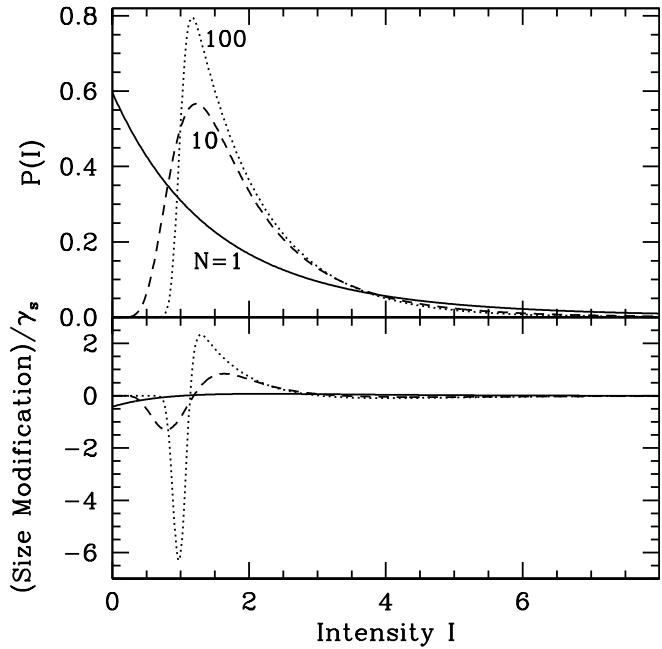}
\caption{
PDF of intensity and the corresponding modification expected for an extended emission region, for $N\!=\!1$, 10, and 100. The source and background intensities are unity, as are all pulse amplitudes $A_j$.}
\label{fig_Size_Compare}
\end{figure}

Of course, the scintillation ensemble must now be included. Integrating each $G_i$ with its respective weight $e^{-G_i}$ gives
\begin{align}
\label{eq::size_1deriv}
&P(I;N,\gamma_{\mathrm{s}}|G) - P(I;N,\gamma_{\mathrm{s}}{=}0|G) \\
\nonumber & \qquad \qquad \approx  2\gamma_{\mathrm{s}} \left( 1 - G\right) \frac{\partial P(I;N,\gamma_{\mathrm{s}}{=}0|G)}{\partial G} .
\end{align}
Here, we have eliminated the degeneracy that appears at linear order between $\gamma_{\mathrm{s},1}$ and $\gamma_{\mathrm{s},2}$ by setting them both equal to a single characteristic size $\gamma_{\mathrm{s}}$.
If the effects of source emission size are substantial, then additional emission information, such as elongation or core/cone geometry, can be inferred from the higher order terms, although higher terms in the original gain expansion must also be retained. To facilitate a comparison with the decorrelation effects derived in \S\ref{sec::Decorr}, we integrate by parts after combining with the final scintillation weight:
\begin{align}
\label{eq::size_residual}
&P(I;N,\gamma_{\mathrm{s}}) - P(I;N,\gamma_{\mathrm{s}} {=} 0) \\
\nonumber & \qquad \approx  - 2\gamma_{\mathrm{s}} \int_0^\infty dG \ G e^{-G} \frac{\partial^2 P(I;N,\gamma_{\mathrm{s}}{=}0|G)}{\partial G^2}.
\end{align}
See Figure \ref{fig_Size_Compare} for a comparison of these effects of emission size for various $N$.

The approximate form of Eq.\ \ref{eq::size_residual} must be applied with some care; it fails in the limit $N \rightarrow \infty$, for example, because of the resulting divergent derivatives of the delta function that characterizes the measured intensity. However, the large-$N$ limit is still easily accessible numerically by applying the full distribution of scintillation gain, as given by Eq.\ 30 in \citet{ISO}. Substituting $\gamma_{\mathrm{s}} = \gamma_{\mathrm{s},1} = \gamma_{\mathrm{s},2}$ in their expression and scaling to normalize the mean gives
\begin{align}
P(\mathcal{G}) = \frac{1+2\gamma_{\mathrm{s}}}{\left(1-\gamma_{\mathrm{s}}\right)^2} &\Biggl[ e^{-(1 + 2\gamma_{\mathrm{s}})\mathcal{G}}\\
\nonumber & - \left( \frac{\mathcal{G}}{\gamma_{\mathrm{s}}} + (1+\mathcal{G}) - 2\mathcal{G}\gamma_{\mathrm{s}} \right ) e^{-\left(2 + \frac{1}{\gamma_{\mathrm{s}}} \right)\mathcal{G}} \Biggr].
\end{align}
When combined with the conditional PDF $P(I;N|\mathcal{G})$ (Eq.\ \ref{eq::I_G_PDF}), this equation produces the scintillation averaged PDF of intensity. The disadvantage of this method is that the PDF must be calculated separately for every value of $\gamma_{\mathrm{s}}$. Fits for emission size effects are then non-linear and computationally expensive, in contrast with the simple linear fit enabled by Eq.\ \ref{eq::size_residual}. Fortunately, the linear expansion of Eq.\ \ref{eq::size_residual} is easily adequate for the full expected range of $N$, even exceeding $10^6$, if $\gamma_{\mathrm{s}} \ll 1$. For $\gamma_{\mathrm{s}} \gsim 0.1$, the linear approximation begins to differ substantially from the exact expression, but this region requires the appropriate quadratic term added to the original gain expansion as well.

Comparison of Eq.\ \ref{eq::size_residual} with Eq.\ \ref{eq::Decorr_1_iid} shows that, at linear order and to the accuracy of the i.i.d.\ approximation for the pulse amplitudes, the effects of an extended emission region on the PDF of intensity are identical to the effects of temporal decorrelation, up to overall scale; this relationship is further analyzed in \S\ref{sec::SizeDecorrCompare}. 
Moreover, the modification arising from extended emission increases with the number of samples averaged, as was the case for temporal decorrelation (see Figure \ref{fig_Size_Compare}). 
These effects also depend on the gated signal-to-noise $S \equiv \left \langle A \right \rangle I_{\mathrm{s}}/I_{\mathrm{n}}$ of the observation; see Figure \ref{fig_Size_SNR_Compare}. Indeed, the dependence on $S$, when combined with the Poisson noise, determines the resolution limit of an observation. We derive this limit explicitly in \S\ref{sec::Expected_Resolution}.

\subsubsection{Effects of a Non-Static Emission Region}
Our analysis is well-suited for long spectral accumulation times, over which the pulsed emission may be highly non-stationary, and the pulsar motion may be substantial. For instance, the emission might be bursty; the received radiation would then represent a superposition of spatially-offset emission sites.
The inferred emission size, as quantified by $\gamma_{\mathrm{s}}$, will then be a characteristic size for the emitting region over the corresponding retarded time. This characteristic size weights the spatial standard deviation by the integrated flux density at each location. 
Thus, spatially-offset emitting regions contribute to the characteristic size, if they emit within the same accumulation time. 
In this sense, our estimate presents an upper limit on emission size, since the emitting region at a single instant may be small relative to the overall emission site. Furthermore, a small emitting region may undergo substantial displacement over the accumulation time, from lateral motion of the pulsar or rotation; both of these effects will present an upward bias of the size estimate.
However, because of the weighting by flux density, weak extended emission may be overwhelmed by a strong pointlike component, yielding a nearly zero characteristic size. Any interpretation of a measured $\gamma_{\mathrm{s}}$ must address these competing factors.

\subsection{Comparison of Emission Size and Temporal Decorrelation}
\label{sec::SizeDecorrCompare}
As we have already observed, comparison of Eq.\ \ref{eq::Decorr_1_iid} with Eq.\ \ref{eq::size_residual} demonstrates that the linear corrections to the intensity PDF for effects of finite emission size and temporal decorrelation are perturbations of identical shape, and differing weights, at least within the excellent i.i.d.\ approximation for the averaged intensities. The relative strength of these two perturbation weights is
\begin{align}
\label{eq::SizeDecorrCompare}
\frac{ \mathrm{size}}{\mathrm{decorrelation}} \approx 24 \frac{\Delta \tau_{\mathrm{d}}^2}{N(N-1)} \gamma_{\mathrm{s}},
\end{align}
where we have substituted the square-law autocorrelation given in \S\ref{sec::Decorr}. Eq.\ \ref{eq::SizeDecorrCompare} allows immediate assessment of the relative importance of effects. For example, if the source extends over $5\%$ of the diffractive scale, then effects arising from the finite emission size will dominate those from decorrelation of the scintillation pattern as long as the averaging is over less than a quarter of the diffractive timescale.

\subsection{Effects of Noise and Bias in Model Parameters}
\label{sec::Obs}

In practice, the estimates of parameters such as the source and background intensity are random variates, and are therefore subject to both noise and bias; we now derive the consequences of such errors. 
These effects, in contrast with effects in the previous sections, do not alter the theoretical PDF of intensity. 
However, they lead to similar changes in the model PDF, which mimic the effects of an extended emission region. They can thus lead to errors in the inferred parameters ($\gamma_{\mathrm{s}}$ and $\Delta \tau_{\mathrm{d}}$) that are distinct from the fundamental limits posed by sampling. 

We have seen that the PDF modifications that arise from both an extended emission region and decorrelation of the scintillation within the averaging time are naturally represented in terms of partial derivatives of $P(I;N|G)$ with respect to $G$, prior to integration with $P(G)$. 
To express all the results of this section in the same way, we again analyze departures using the i.i.d.\ approximation for $P(I;N|G)$.
We also give the expected bias to $\gamma_{\mathrm{s}}$ that arises from each type of parameter error.

\subsubsection{Effects of Biased Background or Source Amplitude}
\label{sec::Biased_Parameters}
Many effects can bias the estimates of the background or source amplitudes. For example, leaked pulse power from quadrature downconversion is frequency-reversed and can be dispersed into the off-pulse region \citep{PaulThesis}. Moreover, the analog-to-digital conversion of the signal requires quantization of the signal; this process introduces an offset of the background noise that varies with signal intensity \citep{Jenet_Anderson,Gwinn_Quant_Noise}.

A bias $\delta I_{\mathrm{n}}$ in the estimated background noise will result in a bias of the estimated pulse amplitudes $\delta A_j = -\delta I_{\mathrm{n}}/I_{\mathrm{s}}$, leading to a bias of the estimated mean intensity within a scintillation element $\delta( \bar{I} \equiv A I_{\mathrm{s}} G + I_{\mathrm{n}} ) = \left(1 - G \right) \delta I_{\mathrm{n}}$. To obtain the incurred PDF modification, we apply the i.i.d.\ approximation and use that $P(I;N|G)$ then only depends on the scale $\bar{I}$ and the degree of averaging $N$. We first expand to leading order in $\delta \bar{I}$, then re-express the derivative using $A I_{\mathrm{s}} \partial/\partial \bar{I} = \partial/\partial G$, and finally integrate $G$ with its exponential weight to obtain
\begin{align}
& P(I;N,\delta I_{\mathrm{n}} {=} 0) - P(I;N,\delta I_{\mathrm{n}})\\
\nonumber & \qquad \approx - \frac{\delta I_{\mathrm{n}}}{A I_{\mathrm{s}}} \int_0^\infty dG\ \left(1 - G \right)e^{-G} \frac{\partial P(I;N,\delta I_{\mathrm{n}}{=}0|G)}{\partial G}.
\end{align}
Comparison with Eq.\ \ref{eq::size_1deriv} shows that this incurred modification is identical to a change in emission size $\gamma_{\mathrm{s}} \rightarrow \gamma_{\mathrm{s}} - \delta I_{\mathrm{n}}/(2 A I_{\mathrm{s}})$. A bias in pulse amplitudes is equivalent, with the substitution $\delta I_{\mathrm{n}} = -I_{\mathrm{s}} \delta A_j$. Thus, at leading order, emission size effects are indistinguishable from a biased estimate of the source or background intensities, regardless of averaging $N$.
However, because the source emission size and the noise bias divided by signal-to-noise likely vary differently with pulse amplitude, segregating the pulses by strength provides a means to resolve this degeneracy.

\subsubsection{Effects of Background or Source Parameter Noise}
\label{sec::Unbiased_Limit}
We similarly analyze the consequence of unbiased noise in the estimated source and background model parameters. This type of parameter error introduces spurious modulation in the corresponding model PDF, which must be compensated by artificial size effects. We therefore expect a positive bias on the inferred size.

Such parameter noise is frequently dominated by the source power estimate because of the limited sampling of the scintillation ensemble in each spectrum. 
This noise $\delta A$ will be effectively independent of $N$, because we have assumed that the scintillation pattern remains approximately constant during the averaging, so each of the $N$ consecutive pulses will be similarly biased. 
The mean intensity within each scintillation element is therefore biased: $\delta \bar{I} = \left( G - 1 \right) I_{\mathrm{s}} \delta A$. However, because this parameter noise is generally unbiased, the leading order correction after an ensemble average is quadratic in $\delta \bar{I}$. Again translating to derivatives with respect to $G$ and averaging over an ensemble of pulse amplitudes, we obtain
\begin{align}
& P(I;N,\delta A {=} 0) -  P(I;N,\delta A)\\
\nonumber & \approx - \frac{1}{2} \left \langle \left( \frac{\delta A}{A} \right)^2 \right \rangle \int_0^\infty dG\ G^2 e^{-G} \frac{\partial^2 P(I;N,\delta A{=}0|G)}{\partial G^2}.
\end{align}
This departure is similar to, though not strictly degenerate with, the effects of a finite emission size. This noise $\delta A$ increases the inferred dimensionless emission size by $\gamma_{\mathrm{s}} \rightarrow \gamma_{\mathrm{s}} + \left \langle \left( \delta A/A \right)^2 \right \rangle/4$. The normalized variance is given by the reciprocal of the number of averaged scintillation elements, $N_{\mathrm{scint}} \sim \mathrm{B}/\Delta \nu_{\mathrm{d}}$, so $\gamma_{\mathrm{s}}$ experiences a positive bias ${\sim} 1/(4 N_{\mathrm{scint}})$. 

The analogous emission size bias from noise in the estimated background intensity is ${\sim} 1/(4 S^2 N_{\mathrm{b}})$, where $S$ is the gated signal-to-noise, and $N_{\mathrm{b}}$ is the number of samples averaged to estimate the background intensity. 
Observe that parameter noise for the background intensity, unlike parameter noise arising from scintillation, is independent for each spectrum and can therefore be mitigated by averaging over series of pulses.

Similarly with effects of bias in the model parameters, the near-degeneracy with effects of finite emission size is broken by comparing multiple analyses. For example, varying the bandwidth used to estimate the model parameters changes their noise; the evolution of measured residual structure with analyzed bandwidth thus precisely quantifies the influence of parameter noise.

\subsection{Effects of Instrumental Distortion }
\label{sec::saturation}
Many instrumental distortions depend on the signal amplitude; they can therefore be modeled by imposing appropriate distortions on the distribution of scintillation gain $P(G)$. 
The influence on the PDF can be modeled by combining the modified $P(G)$ with the conditional distributions $P(I;N|G)$ derived above. 
Alternatively, the modulation index $m$ (Eq.\ \ref{eq::modulation_index_simple}) provides a simple measure to quantify changes in $P(G)$; we therefore use $m$ to characterize effects in this section.

Instrumental distortions can arise, for example, from saturation of the observing system.
Because strong pulsars can be highly variable and dominate backgrounds when they are on, optimal quantization levels are often difficult to determine, and even systems with many bits may saturate. A simplified model of saturation imposes a voltage cutoff on the observed scalar electric field in the time-domain: 
\begin{align}
\hat{x}_i = 
\left\{
  \begin{array}{l l}
    x_i & \quad \text{if $|x_i| < \sigma_{\mathrm{max}}$},\\
    \sigma_{\mathrm{max}}\mathrm{sign}(x_i) & \quad \text{else.}\\
  \end{array} \right.
\end{align}

The effect of this cutoff is two-fold: power from the strongest samples is spread across all channels in the spectral domain, and the observed signal intensity is underestimated. At leading order, the second effect predominates, so that the estimated off-pulse noise overestimates the effective on-pulse noise. In terms of the discussion in \S\ref{sec::Biased_Parameters}, saturation leads to a bias $\delta I_{\mathrm{n}} < 0$ and thus, an inferred emission size that is upward biased.

Likewise, if the pulse is heavily dispersed, then the saturation predominantly affects samples with the highest scintillation gain $G$. As a simplified model of this effect, we replace the distribution of scintillation gain $P(G) = e^{-G}$ by a truncated version: $\hat{P}(G) \equiv e^{G}\theta\left(G_{\mathrm{max}} - G \right) + e^{-G_{\mathrm{max}}} \delta\left(G-G_{\mathrm{max}} \right)$, where $\theta(x)$ is the Heaviside step function. The corresponding modulation index is
\begin{align}
\label{eq::saturation_modulation}
\hat{m}^2 = \frac{\left \langle G^2 \right \rangle_{\hat{P}}}{\left \langle G \right \rangle_{\hat{P}}^2} - 1 = \frac{\sinh(G_{\mathrm{max}}) - G_{\mathrm{max}}}{\cosh(G_{\mathrm{max}}) - 1} \leq 1.
\end{align}
Comparison with Eq.\ \ref{eq::modulation_index_simple} shows that even a relatively large cutoff $G_{\mathrm{max}} = 5$ gives suppression equivalent to modest emission size effects, $\gamma_{\mathrm{s}} \approx 0.014$. 
Furthermore, the exponential character of the incurred suppression leads to an extremely rapid onset of saturation artifacts once the largest intensities clear the threshold for saturation; the effects on the strongest pulses may therefore be quite pronounced relative to those on the weaker pulses.

\subsection{Expected Resolving Power}
\label{sec::Expected_Resolution}
\begin{figure}[t]
\includegraphics*[width=0.48\textwidth,clip=true]{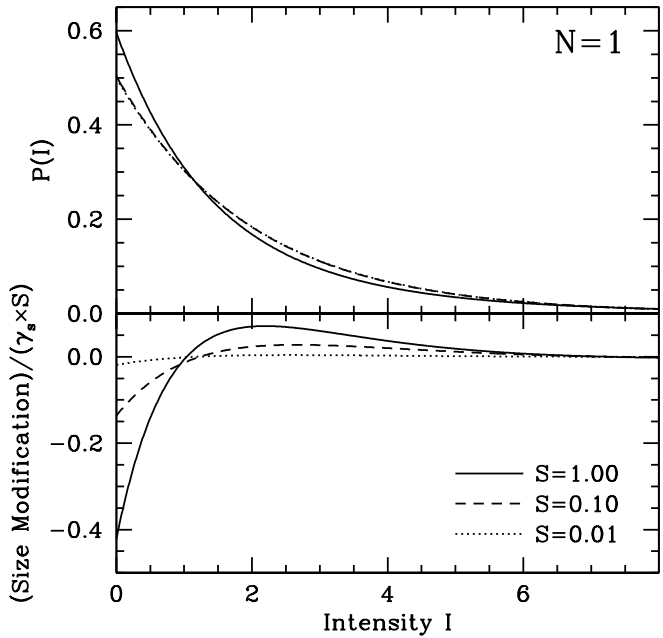}
\caption{
Intensity PDF and the corresponding modification for emission size effects, for $N\!=\!1$ and varying gated signal-to-noise ($S$). In each case, $\langle A \rangle I_{\mathrm{s}} + I_{\mathrm{n}} = 2.0$. Observe the the plotted modification is scaled by both the dimensionless size parameter ($\gamma_{\mathrm{s}}$) and $S$. 
}
\label{fig_Size_SNR_Compare}
\end{figure}

Poisson noise presents a fundamental limit on the resolving power afforded by scintillation, as a fraction of the magnified diffractive scale. 
This limit depends on both the gated signal-to-noise ratio $S \equiv \left\langle A \right \rangle I_{\mathrm{s}}/I_{\mathrm{n}}$ and the observational parameters $B$ and $T_{\mathrm{obs}}$. Figure \ref{fig_Size_SNR_Compare} illustrates the variation of PDF modification with $S$, for $N=1$. The modification from extended emission size is maximal at zero intensity, with fractional strength
\begin{align}
\label{eq::ResidualStrength}
&\left|\frac{P(I{=}0;N{=}1,\gamma_{\mathrm{s}}) - P(I{=}0;N{=}1,\gamma_{\mathrm{s}}{=}0)}{P(I{=}0;N{=}1,\gamma_{\mathrm{s}}{=}0)} \right| \\
\nonumber &\qquad \qquad = -2 \gamma_{\mathrm{s}} \left[ 2 + \frac{1}{S} + \left(1+S\right) \frac{e^{-1/S}}{\mathrm{Ei}\left(-1/S \right)} \right], 
\end{align}
where $\mathrm{Ei}(x)$ is the imaginary error function. Equating Eq.\ \ref{eq::ResidualStrength} with the level of Poisson noise $\eta$ in the observed PDF then provides an excellent assessment of resolving power.

We apply an approximation for Eq.\ \ref{eq::ResidualStrength} to obtain the following estimate for the achievable resolution, $\sqrt{\gamma_{\mathrm{s},\mathrm{min}}}$, in units of the magnified diffractive scale:
\begin{align}
\sqrt{\gamma_{\mathrm{s},\mathrm{min}}} \approx \sqrt{\frac{\eta}{S}} \exp\left[0.2 + 0.09 \left(\ln S\right)^2 + 0.005 \left(\ln S \right)^3 \right].
\end{align}
This approximation is good to within $10\%$ for $10^{-3} < S < 2$. Simulations indicate that substituting $\eta \equiv \frac{1.3}{\sqrt{N_{\mathrm{tot}}}}$ effectively estimates the standard errors from the full distribution fits, where $N_{\mathrm{tot}}$ is the total number of sampled points used to estimate the PDF of intensity.

We now apply this result to estimate the resolving power for typical observational parameters. 
A $(1\mathrm{\ hour})\!\times\!(4\mathrm{\ MHz})$ observation of a pulsar with $5\%$ duty cycle gives ${\sim} 10^9$ on-pulse samples of the flux density. A gated SNR $S=1$ then gives a $3\sigma$ detection limit of about $1\%$ of the magnified diffractive scale, $S=0.1$ gives about $10\%$ of the magnified diffractive scale, and $S=0.01$ gives about $50\%$ of the magnified diffractive scale. Since the detection limit is much more sensitive to $S$ than to the number of sampled points, an analysis of a subset of strong pulses may provide optimal resolution.

\section{Moments of the Distributions: The Modulation Index and Self-Noise}  
\label{sec::Moments}

Moments of the intensity distribution provide simple diagnostics for scattering and source inference. Frequently, moments provide optimal estimators of distribution parameters and are therefore both powerful and elegant tools for characterizing distributions. However, a blind calculation of moments may be blind to effects of RFI, artifacts of finite scintillation averages, or deficiencies of models, whereas a comparison with the full distribution functions might identify such effects.

We focus on two applications of moments of the distribution: the intensity modulation index and a practical technique for estimating self-noise. For the modulation index, we include all sources of noise, as well as evolution of the scintillation pattern and spatially-extended source emission. For the self-noise estimator, we constrain amplitude variations of the pulsar and assume that there is no decorrelation of the scintillation pattern within the averaging time ($N \ll \Delta \tau_{\mathrm{d}}$); we then assess departures from these assumptions qualitatively.

\subsection{The Modulation Index}
\label{sec::Modulation_Index}
In \S\ref{sec::Size}, we demonstrated that the size of the emission region affects the modulation index $m$. 
However, self-noise, temporal decorrelation, and background noise also affect the modulation index.
These effects are readily evaluated, either by analysis of the random variables that make up each sample of intensity (Eq.\ \ref{eq::E_I}), or by appropriate derivatives of the corresponding characteristic function. 

Here, we analyze only the case for which the pulse amplitudes are i.i.d.: $\left \langle A_i A_j \right \rangle = \left \langle A \right \rangle^2 + \delta_{ij} \left \langle \delta A^2 \right \rangle$. This condition is different than the assumptions for the i.i.d.\ approximation derived above, which assumes the stronger condition that the spectral statistics are i.i.d.. Although many pulsars appear to have i.i.d.\ pulse amplitudes, others display striking departures. For example, phenomena such as nulling and mode-changing break statistical isotropy, and the brightest pulses from the Vela pulsar tend to occur in groups \citep{Palfreyman}.

To explicitly account for the temporal decorrelation of the scintillation pattern, we again assign the autocorrelation appropriate for square-law scattering, quantified by the decorrelation timescale (in pulse periods) $\Delta \tau_{\mathrm{d}}$. 
We incorporate effects of an extended emission region in terms of the dimensionless size parameter $\gamma_{\mathrm{s}}$, defined by Eq.\ \ref{eq::gamma_s}, with the assumption that $\gamma_{\mathrm{s}} \ll 1$. Modification of the scattering assumptions will introduce minor scaling alterations for these contributions.

To leading order in the effects of emission size and temporal decorrelation, the modulation index is then given by
\begin{align}
\label{eq::mod_I}
m^2 &\equiv \frac{ \left \langle (I - I_{\mathrm{n}})^2 \right \rangle }{ \left \langle I - I_{\mathrm{n}} \right \rangle^2} - 1 \\
\nonumber &\approx  \ 1 -  \overbrace{4 \gamma_{\mathrm{s}} \left( 1 + \frac{1}{N} \right)}^{\mbox{\hspace{-1cm}\small{Emission Size\hspace{-1cm}}}}
\ +\  {\overbrace{\frac{2}{N}}^{\mbox{\hspace{-1cm}\small{Self-Noise}\hspace{-1cm}}}}
\ +\  \overbrace{\frac{1}{N S}\left( 2 + \frac{1}{S} \right)}^{\mbox{\hspace{-1cm}\small{Signal-to-Noise}\hspace{-1cm}}}\\
\nonumber &\quad  + \underbrace{4 \left(1 - 2\gamma_{\mathrm{s}} \right) \frac{\left \langle \delta A^2 \right \rangle}{N \left \langle A \right \rangle^2}}_{\mbox{\hspace{-1cm}\small{Intrinsic Variability}\hspace{-1cm}}} 
\ -\  \underbrace{\left( 1 - 4 \gamma_{\mathrm{s}} \right) \frac{N^2-1}{6 \Delta \tau_{\mathrm{d}}^2}}_{\mbox{\hspace{-1cm}\small{Decorrelation}\hspace{-1cm}}}.
\end{align}
Here, we have grouped the contributions to the modulation index, to explicitly illustrate the relative influence of the various effects. We also have used $\gamma_{\mathrm{s},1} = \gamma_{\mathrm{s},2} \equiv \gamma_{\mathrm{s}}$ because of the degeneracy at linear order. We see that the relative contributions of an extended emission region and decorrelation of the scintillation pattern give the previously derived relationship between their respective PDF modifications (Eq.\ \ref{sec::SizeDecorrCompare}). 

The modulation index corresponds to the zero time and frequency lags in the autocorrelation function of intensity; analysis of non-zero lags eliminates many of the terms in Eq.\ \ref{eq::mod_I}. For example, non-zero spectral lags eliminate the `Signal-to-Noise' contribution and mitigate the `Self-Noise' contribution (although intrinsic modulation induces correlations in the spectral self-noise that preserve the contribution). Non-zero temporal lags fully eliminate all terms other than `Emission Size' and `Decorrelation,' which is enhanced.

\subsection{An Estimate of Self-Noise}
\label{sec::SelfNoise}
The self-noise of a signal is a useful diagnostic, which can be used to identify intrinsic variability on timescales shorter than the spectral accumulation time. Intrinsic modulation induces correlations among frequencies, without modifying the mean spectrum \citep{Intermittent_Noiselike_Emission}. If spectra are formed according to the criteria in \S\ref{sec::instrumental}, then the power in each spectral channel is an exponential random variable. Because such variables are completely characterized by their mean, the single-channel statistics are therefore immune to intrinsic modulation. However, the intrinsic modulation does introduce correlations in the noise of nearby channels. 
We now derive the expected self-noise in the spectral domain, thereby enabling detection of such correlations.

One method for estimating self-noise is to compare pairs of nearby samples that are assumed to be within the same scintillation element. We will assume that the pair consists of samples with uncorrelated self-noise (e.g.\ pairs from different pulses, or from the same pulse with negligible intrinsic modulation). Because self-noise is heteroscedastic (see \S\ref{sec::GaussianApproximation}), we calculate the noise $\delta I^2 \equiv \left(I_1 - I_2\right)^2\!/2$ in pairs of samples $\{I_1,I_2\}$ as a function of their mean $I = (I_1 + I_2)/2$; we denote the resulting noise estimate $\delta I^2(I)$. The ensemble average is given by
\begin{align}
\delta I^2(I) = \frac{\int_0^{2I} 2(I - I')^2 P(I', 2 I - I';N) dI'}{\int_0^{2I} P(I', 2 I - I';N) dI'},
\end{align}
where $P(I_1,I_2;N)$ represents the probability of sampling the pair $\{I_1,I_2\}$, including an ensemble average over scintillation. Because we assume that the samples are independent but are drawn from the same scintillation element, $P(I_1,I_2;N) = \int_0^\infty dG\ P(I_1;N|G) P(I_2;N|G)$.

As usual, we apply the i.i.d.\ approximation. Although the scintillation integral cannot be evaluated in closed form, Eq.\ \ref{eq::I_iid} readily yields the identity
\begin{align}
\label{eq::SelfNoiseIdentity}
& \int_0^{2I} 2(I - I')^2 P(I', 2 I - I';N) dI'\\
\nonumber &\qquad \qquad = \frac{I^2}{N + \frac{1}{2}} \int_0^{2I} P(I', 2 I - I';N) dI'.
\end{align}
This relationship immediately unveils the simple relationship between the estimated mean signal and noise:
\begin{align}
\delta I^2(I) = \frac{I^2}{N + \frac{1}{2}}.
\end{align}
In fact, because the identity given by Eq.\ \ref{eq::SelfNoiseIdentity} holds regardless of the integration over $G$, 
this property of the estimated self-noise is independent of the degree of sampling and the properties of scintillation.
We see that this method for estimating the self-noise gives the same form as the exact expression (see \S\ref{sec::GaussianApproximation}), except that $N \rightarrow N + 1/2$.

If the averaged intensities are from different pulses, then pulse-to-pulse variations contribute additional noise. If the averaged intensities are from the same pulses, then intrinsic variations on timescales shorter than $t_{\mathrm{acc}}$ induce correlations in self-noise, and thereby decrease the measured noise; if $N>1$ then pulse-to-pulse variations within the averaging will increase the noise. Each of these effects can be estimated by applying the appropriate substitution $N\rightarrow N_{\mathrm{eff}}$. 
Similar tests were applied by \citet{Noise_Inventory} to infer short-timescale variability ($<300 \mu s$) of PSR B0834+06.

\section{Summary}
\label{sec::Summary}
We have presented observable quantities that describe the flux density of a source exhibiting strong diffractive scintillation. These observables include both distribution functions and bulk indicators, such as the modulation index and estimates of self-noise. Our results encompass a broad range of physical and instrumental effects, such as non-stationary background noise, arbitrary temporal averaging, pulse-to-pulse variability, decorrelation of the scintillation pattern within the averaging, and the possibility of spatially-extended source emission, making them well-suited for direct comparison with observations. The primary benefits of such comparisons include the following:
\begin{itemize}
\item We can completely decouple effects arising from decorrelation of the scintillation pattern from those of extended emission size by analyzing Nyquist-limited data without averaging.
\item Our results enable estimation of the emission region sizes of individual pulses and of arbitrary subclasses of pulses.
\item For point-source emission, our statistical description does not require specification of either the nature or geometry of the scattering material. Residuals relative to the point-source model of \S\ref{sec::PDist} therefore provide robust indicators of either intrinsic emission effects (e.g.\ finite size of the emission region), extrinsic effects (e.g.\ evolution of the scintillation pattern), or instrumental limitations (e.g.\ finite observing bandwidth). These competing effects can be distinguished by varying the degree of averaging, the analyzed bandwidth or the chosen subset of pulses.
\item Our technique does not appeal to extraordinary scattering geometry (such as is inferred for parabolic arcs) or extreme scattering events to infer the size of the emission region. Furthermore, our analysis is effective on short observations ($\lsim\! 1$ hour) and is easily reproducible. The achievable resolution limits (see \S\ref{sec::Expected_Resolution}) can be a small fraction of the magnified diffractive scale.
\end{itemize}

In particular, our results are ideal for describing the statistics of low-frequency observations of scintillating pulsars. In such cases, the scintillation bandwidth can be narrow relative to the observing bandwidth, allowing an estimate of the scintillation-averaged source intensity for each pulse. The tight coupling of scintillation and source variations is then described by our models, which fully incorporate the intrinsic source amplitudes and the background noise estimates via on-pulse and off-pulse averages. 
As we report elsewhere, we have found excellent agreement between these theoretical expectations and observations of the Vela pulsar at $800$ MHz \citep{GUPPI_size_2}. We thereby achieved a spatial resolution of ${\sim}1\%$ of the magnified diffractive scale -- about 4 km at the pulsar. Application to other pulsars will enrich the empirical description of pulsar radio emission regions, potentially including variation among individual pulses and pulse subclasses.

\acknowledgments
We thank the U.S. National Science Foundation for financial support for this work (AST-1008865).

\appendix

\section{Mathematical Results}
\subsection{Average of Exponential Random Variables with Different Scales}
Consider an exponential random variable $I$: $P(I;\bar{I}) = \bar{I}^{-1} e^{-I/\bar{I}}$ for $I>0$, where $\bar{I} \equiv \left \langle I \right \rangle$. The corresponding characteristic function is $\varphi(k) = \left(1 - i \bar{I} k\right)^{-1}$. The distribution $P\left(I;\{\bar{I}_j\}\right)$ of the average of $N$ such variables from independent distributions with different scales $\{\bar{I}_j\}$ is then easily derived by inverting the partial fraction decomposition of the product of characteristic functions:
\begin{align}
\label{eq::mult_I}
P\left(I;\{\bar{I}_j\}\right) = N \sum_{j=1}^N \left( \frac{\bar{I}_j^{N-2} }{\prod\limits_{\substack{\ell = 1\\ \ell\neq j}}^N \left(\bar{I}_j - \bar{I}_\ell \right)}  \right) e^{-N I/\bar{I}_j},\quad I>0.
\end{align}
In other words, the distribution of the average of exponential random variables is a weighted sum of the marginal distributions. This sum is the hypoexponential (or generalized Erlang) distribution.

The representation of Eq.\ \ref{eq::mult_I} is singular if any of the scales are equal. In particular, if the random variables are i.i.d., each with scale $\bar{I}$, then their average is drawn from an Erlang distribution, which is simply a gamma distribution with integer shape parameter:
\begin{align}
\label{eq::I_iid}
P\left(I;\bar{I},N\right) = \frac{N^N}{(N-1)!}  \frac{I^{N-1}}{\bar{I}^N}  e^{-\frac{N I}{\bar{I}}},\quad I>0.
\end{align}
\citet{Goodman_SO} discusses both these distributions in the similar context of polarized thermal light.

\subsection{Distribution of Correlated Exponential Random Variables}
\label{sec::multI}
We now approximate the distribution of $N$ correlated exponential random variables $\{I_i\}$. The exact distribution follows by first constructing a complex multivariate normal distribution with the desired covariance matrix; the exponential random variables are the squared norms of each circular complex Gaussian random variable. We restrict ourselves to the case in which each random variable has unit mean. The characteristic function of $P\left(\{ I_i \}\right)$ can then be written \citep{mallik}
\begin{align}
\label{eq::char_multI}
\varphi\left( \{ k_i \}; \boldsymbol{\Sigma} \right) = \frac{1}{\mathrm{det}\left[ \textbf{I} - i \mathrm{diag}\left\{ k_i \right\} \boldsymbol{\Sigma} \right]}.
\end{align}
Here, $ [\boldsymbol{\Sigma}]_{ij} = \sqrt{ \langle I_i I_j \rangle - 1 } \equiv \sqrt{\Gamma_{ij}}$. 

We now expand in $\boldsymbol{\epsilon} \equiv \textbf{J} - \boldsymbol{\Sigma}$, where $\textbf{J}$ is the $N\!\times\! N$ unit matrix: $[\textbf{J}]_{ij} = 1\ \forall\ i,j$. Since $[\boldsymbol{\epsilon}]_{ij} \approx \left( 1 - \Gamma_{ij} \right)/2$, we can approximate the characteristic function to leading order as
\begin{align}
\varphi\left( \{ k_i \}; \boldsymbol{\Sigma} \right) = \frac{1}{1 - i \sum_{j=1}^N k_j} + \frac{\sum_{i<j} k_i k_j \left( 1 - \Gamma_{ij}\right)}{\left(1 - i \sum_{j=1}^N k_j\right)^2} + \mathcal{O}\left( \left[ 1 - \Gamma_{ij} \right]^2 \right).
\end{align}
The inverse Fourier transform of $\varphi\left( \{ k_i \}; \boldsymbol{\Sigma} \right)$ then gives our approximation to $P(\{ I_i \})$. Although this inverse can be written in terms of delta functions and derivatives, the most natural form expresses the action as a weight function:
\begin{align}
\label{eq::multI_approx}
\int d^N I_i \ f(\{ I_i \}) P(\{ I_i \}) = \int_0^\infty dI \ \left[ f(\{I\}) e^{-I} - \sum_{i<j} \left(1 - \Gamma_{ij} \right) I e^{-I} \left.\frac{\partial^2 f}{\partial I_i \partial I_j}\right\rfloor_I + \mathcal{O}\left( \left[ 1- \Gamma_{ij} \right]^2 \right) \right].
\end{align}
Here $f(\{I\})$ denotes $f(\{I_i = I\ \forall \ i\})$.

\subsubsection{Example: Bivariate PDF of Correlated Exponential Random Variables}
The joint distribution of a pair of correlated exponential random variables is particularly simple. In this case, the characteristic function (\ref{eq::char_multI}) is easily inverted:
\begin{align}
\label{eq::I2}
P(I_1,I_2;\Gamma) = \frac{1}{1-\Gamma} I_0\left( \frac{2\sqrt{\Gamma}}{(1-\Gamma)} \sqrt{I_1 I_2} \right) \exp\left[ -\frac{1}{(1-\Gamma)} \left(I_1 + I_2 \right) \right],\quad I_1,I_2>0.
\end{align}
Here, $\Gamma = \langle I_1 I_2 \rangle - 1$, and $ I_0(x)$ is the modified Bessel function. Using Eq.\ \ref{eq::multI_approx}, we can approximate Eq.\ \ref{eq::I2} as
\begin{align}
\label{eq::I2_approx}
P(I_1,I_2;\Gamma) = e^{-\hat{I}} \left\{ \delta(\Delta I) + \left(1 - \Gamma\right) \left[ \frac{2 - \hat{I}}{4} \delta(\Delta I) + \hat{I} \delta''(\Delta I) \right]  + \mathcal{O}\left[\left(1 - \Gamma \right)^2 \right]\right\},\quad \hat{I}>0,
\end{align}
where $\hat{I} \equiv (I_1 + I_2)/2$ and $\Delta I \equiv I_1 - I_2$.

\subsubsection{Convolution of Correlated Exponential Random Variables}
\label{sec::ConvolutionCorrelated}
The gamma distribution is an excellent approximation for the sum of independent exponential random variables that are not necessarily isotropic (see \S\ref{I_PDF_iid}) and can also be used to approximate the convolution of correlated exponential random variables. This approximation is most effective for correlations that are particularly strong or weak. For example, consider the average of two exponential random variables, each with unit mean, and with covariance $\left \langle I_1 I_2 \right \rangle = 1 + \Gamma$.
The gamma distribution that matches the mean and variance of the exact average has a corresponding effective number of degrees of freedom $N_{\mathrm{eff}} = 2/(1+\Gamma)$. These distributions agree exactly for the limits $\Gamma \rightarrow 0,1$, but suffer sizeable discrepancies for intermediate correlations. For example, the third central moment of the average of two correlated exponential random variables is $(1 + 3 \Gamma )/2$ but is $2/N_{\mathrm{eff}}^2 = (1 + \Gamma )^2/2$ for the corresponding gamma distribution -- a difference of ${\sim} 10\%$ for intermediate correlation. 

More generally, for $N$ averaged exponential random variables with correlation matrix $\Gamma_{ij}$, as defined above, and unit mean, the appropriate number of degrees of freedom based on variance matching is $N_{\mathrm{eff}} = N\left(1 + 2N^{-1} \sum_{i<j} \Gamma_{ij}\right)^{-1}\!$. Even in the case of large $N$, the disparity in skewness between the exact average and the appropriate gamma distribution remains. For instance, if $\Gamma_{ij} \equiv \Gamma$ for $i\neq j$ then,
\begin{align}
\mu_3 = 
\begin{cases} 2\Gamma^{3/2} + \frac{6\Gamma}{N} \left(1 - \sqrt{\Gamma} \right) + \frac{2}{N^2} \left(1 - 3 \Gamma + 2 \Gamma^{3/2} \right) & \text{Exact Average,}
\\
2\Gamma^2 + \frac{4\Gamma}{N} \left(1 - \Gamma \right) + \frac{2}{N^2}\left(1-\Gamma\right)^2 &\text{Approximation with }N_{\mathrm{eff}} = \frac{N}{1 + (N-1)\Gamma}\text{,}
\end{cases}
\end{align}
where $\mu_3$ denotes the third central moment. Again, we see that these expressions agree for $\Gamma \rightarrow 0,1$ as expected, but differ substantially for intermediate correlation, even as $N\rightarrow \infty$.\\ 

\bibliography{High_Resolution_Intensity_Statistics_References.bib}

\end{document}